# Change with energy of source geometry as seen in hadron interferometry[*]

I.V. Andreev[1,2†], M. Plümer[1‡], B.R. Schlei[1§] and R.M. Weiner[1¶]

[1] Physics Department, Univ. of Marburg, Marburg, FRG
[2] P.N. Lebedev Institute, Moscow, Russia

**Abstract**

The variation of radii, lifetimes, correlation lengths and the chaoticity parameter of particle sources with the collision energy $\sqrt{s}$ is studied by analysing and comparing two-particle Bose-Einstein correlation data obtained by the NA22 Collaboration at $\sqrt{s} = 22\ GeV$ and by the UA1-Minimum-Bias Collaboration at $\sqrt{s} = 630\ GeV$. The UA1-data are found to be inconsistent with a static source, whereas both the data at $\sqrt{s} = 22\ GeV$ and at $630\ GeV$ can be described in an expanding source model. In the latter model, the $s$-dependence of the data implies an increase of the transverse radius with $s$ and a decrease of the correlation lengths in longitudinal and in transverse direction. The chaoticity parameter increases or remains approximately constant as a function of the CM energy.

The study of the energy dependence of physical quantities has been an important tool in multiparticle dynamics. Thus e.g. the discovery of the violation of KNO scaling with the increase of energy from ISR to SPS energies has been one of the most stimulating findings in soft strong interactions in the last decade.

While it is now understood that this phenomenon is due to the energy dependence of long range correlations [1] it was unclear whether short range correlations (SRC) do also depend on energy. Besides its heuristic interest this question is also of practical interest since it is related to the energy dependence of radii and lifetimes which determine via Bose-Einstein correlations (BEC) the SRC of identical particles.

Here we present the first results of an investigation of the energy dependence of source parameters in BEC. These results were made possible by

- the publication of BEC data at two quite different values of $\sqrt{s}$, i.e. $\sqrt{s} = 22 GeV$ [2] and $\sqrt{s} = 630 GeV$ [3]

- the development of a space-time theory of BEC [4].

Two models were considered for this analysis (for details on these models cf. [5]).

1. A static source defined by the source distributions $f_{ch}$ (for the chaotic part), $f_c$ (for the coherent part) and the correlation $C$:

$$f_{ch}(x) \propto \exp\left(-\frac{x_0^2}{R_{ch,0}^2} - \frac{x_\parallel^2}{R_{ch,\parallel}^2} - \frac{x_\perp^2}{R_{ch,\perp}^2}\right) \quad (1)$$

$$f_c(x) \propto \exp\left(-\frac{x_0^2}{R_{c,0}^2} - \frac{x_\parallel^2}{R_{c,\parallel}^2} - \frac{x_\perp^2}{R_{c,\perp}^2}\right) \quad (2)$$

and

$$C(x - x') \propto \exp\left[-\frac{(x_0 - x_0')^2}{2L_0^2} - \frac{(x_\parallel - x_\parallel')^2}{2L_\parallel^2} - \frac{(\vec{x}_\perp - \vec{x}_\perp')^2}{2L_\perp^2}\right] \quad (3)$$

where $R_{ch,\alpha}$ and $R_{c,\alpha}(\alpha = 0, \perp, \parallel)$ are the lifetimes, transverse radii and longitudinal radii of the chaotic source and of the coherent source, respectively, and where $L_\alpha(\alpha = 0, \perp, \parallel)$ are the correlation time and the corresponding correlation lengths in transverse and in longitudinal direction. To specify the relative contributions of the chaotic and the coherent component, one needs to fix the value of the (momentum dependent) chaoticity parameter $p(k^\mu)$ at some arbitrary scale, e.g., $p_0 \equiv p(k^\mu = 0)$.

2. A longitudinally expanding source

$$f_{ch}(x) \propto \delta(\tau - \tau_0) \exp\left(-\frac{x_\perp^2}{R_{ch,\perp}^2}\right) \quad (4)$$

$$f_c(x) \propto \delta(\tau - \tau_0) \exp\left(-\frac{x_\perp^2}{R_{c,\perp}^2}\right) \quad (5)$$

$$C(\eta - \eta', x_\perp - x_\perp') \propto \exp\left[-\frac{2\tau_0^2}{L_\eta^2}\sinh^2\left(\frac{\eta - \eta'}{2}\right) - \frac{(\vec{x}_\perp - \vec{x}_\perp')^2}{2L_\perp^2}\right] \quad (6)$$

$\tau$ is the longitudinal proper time and $\eta$ the space-time rapidity.

The number of free parameters can be reduced by relating them to the shape of the one-particle inclusive distribution $\rho_1(\vec{k})$. For the static source, one has

$$\begin{aligned}\rho_1(\vec{k}) &= N\left[p_0 \exp\left(-\frac{1}{2}(E^2 R_{L0}^2 + k_\parallel^2 R_{L\parallel}^2 + k_\perp^2 R_{L\perp}^2)\right)\right. \\ &\quad \left. + (1 - p_0) \exp\left(-\frac{1}{2}(E^2 R_{c,0}^2 + k_\parallel^2 R_{c,\parallel}^2 + k_\perp^2 R_{c,\perp}^2)\right)\right]\end{aligned} \quad (7)$$

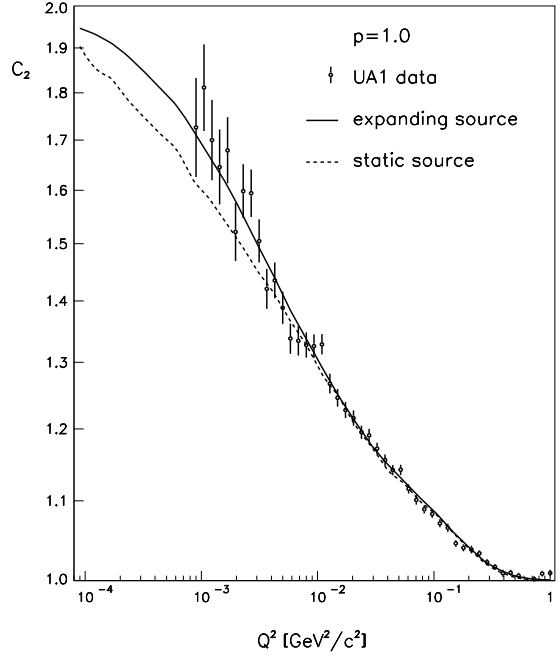

Figure 1: Best fits to the UA1 data

and for the expanding source,

$$\rho_1(\vec{k}) = N \frac{m}{m_\perp} \left[ p_0 \exp\left(-\frac{1}{2} k_\perp^2 R_{L\perp}^2\right) + (1 - p_0) \exp\left(-\frac{1}{2} k_\perp^2 R_{c,\perp}^2\right) \right] \quad (8)$$

where $N$ is a normalization factor and

$$R_{L\alpha}^2 \equiv \frac{R_{ch,\alpha}^2 L_\alpha^2}{R_{ch,\alpha}^2 + L_\alpha^2} \qquad (\alpha = 0, ||, \perp). \quad (9)$$

For the static source, a plateau in rapidity requires $R_{c,0} = R_{c,||} = L_0 = L_{||} = 0$. Furthermore we make the simplifying assumption that the particle spectra have the same shape for the chaotic and for the coherent component, i.e., $R_{L\perp} = R_{c,\perp}$. This implies that the fraction of chaotically produced particles is momentum independent. Therefore, in what follows we have dropped the subscript "0" in the notation for the chaoticity parameter. Finally, the value of $R_{L\perp} = R_{c,\perp}$ can be determined by requiring the average transverse momentum $\langle k_\perp \rangle$ as calculated from Eqs. (7,8) to agree with the experimentally observed value[1].

Thus, the remaining four free parameters are $R_{ch,0}$, $R_{ch,||}$, $R_{ch,\perp}$ and $p$ for the static source, and $\tau_0$, $R_{ch,\perp}$, $L_\eta$ and $p$ for the expanding source.

---

[1] A comparison of NA22- [6] and Tevatron-data [7] shows that the average transverse momentum *of pions* depends only weakly on the CM energy. In our calculations, we have taken $\langle k_\perp \rangle \sim 0.37\, GeV$.

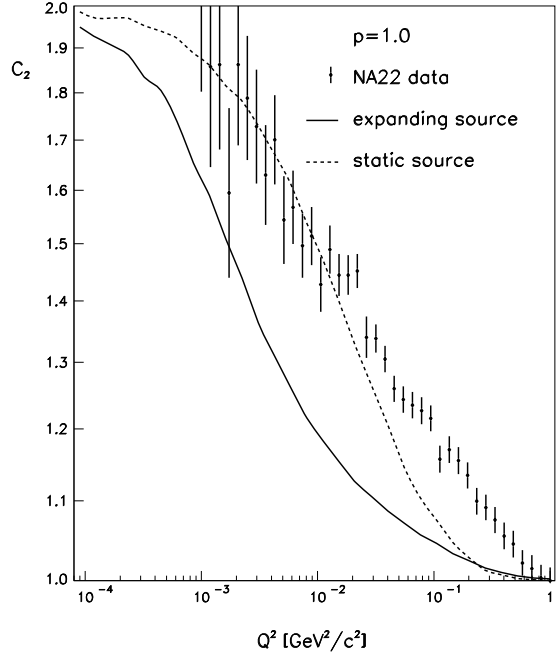

Figure 2: Curves for the NA22 phase space cuts, obtained for the same parameter values as in Fig. 1a.

Since the data are expressed in terms of the invariant variable $Q^2$, one has to calculate the correlation function in terms of this variable. The correlation function reads[2]:

$$\tilde{C}_2(Q^2) = \frac{\int_\Omega d\omega_1 \int_\Omega d\omega_2 \ \rho_2(\vec{k}_1, \vec{k}_2) \ \delta\left[Q^2 + (k_1^\mu - k_2^\mu)^2\right]}{\int_\Omega d\omega_1 \int_\Omega d\omega_2 \ \rho_1(\vec{k}_1)\rho_1(\vec{k}_2) \ \delta\left[Q^2 + (k_1^\mu - k_2^\mu)^2\right]} \qquad (10)$$

Below we discuss our results:

Fig. 1 shows the UA1-data[3] for the two-particle BEC as a function of $Q^2$ together with the fits obtained in refs. [8, 9] for a purely chaotic source ($p = 1$), for the static and the expanding source, respectively. The corresponding parameter values and the values of $\chi^2/d.o.f.$ are listed in Tables 1 and 2.

---

[2] It is worth mentioning that the same type of integrals occur in the interpretation of "intermittency" data. In ref. [8, 9] it was shown that a conventional source of the type defined above with a fixed geometry (at a given energy) can explain all the so-called intermittency observations, i.e. the dependence of moments on $Q$ and there is no need for a fractal source as suggested in [10]. As a matter of fact a fractal source does not imply even a simpler explanation of the data because it involves an exotic length scale of the order of 6-7 fm for which there is no simple interpretation

[3] We have normalized both the UA1-data [3] and the NA22-data [2] to unity at $Q^2 = 1 \ GeV^2$ in order to subtract the effects of long range correlations which were shown to play an important role at UA1-energies[1].

| Static source | | | | | |
|---|---|---|---|---|---|
| $\sqrt{s}$ [GeV] | $p$ | $R_{ch,0}$ [fm] | $R_{ch,\|}$ [fm] | $R_{ch,\perp}$ [fm] | $\chi^2/d.o.f.$ |
| 630 | 1.00 | 1.00 | 3.50 | 1.05 | 1.65 |
| 22 | 1.00 | 3.80 | 0.50 | 1.00 | 1.42 |
| 22 | 0.60 | 3.00 | 1.00 | 1.00 | 1.00 |

Table 1: Fit parameter values and $\chi^2/d.o.f.$ for the static source.

| Expanding source | | | | | |
|---|---|---|---|---|---|
| $\sqrt{s}$ [GeV] | $p$ | $\tau_0$ [fm] | $R_{ch,\perp}$ [fm] | $L_\eta$ [fm] | $\chi^2/d.o.f.$ |
| 630 | 1.00 | 2.00 | 1.00 | 0.40 | 1.33 |
| 22 | 1.00 | 1.90 | 0.65 | 0.90 | 0.99 |
| 22 | 0.60 | 1.90 | 0.70 | 0.90 | 0.90 |

Table 2: Fit parameter values and $\chi^2/d.o.f.$ for the expanding source.

The NA22-data are plotted in Fig. 2. Clearly, the BEC function $\tilde{C}_2(Q^2)$ at $\sqrt{s} = 22\ GeV$ is broader than the one at 630 $GeV$. Naively, one might conclude from this observation that the radii and/or lifetimes of the source increase with CM energy. However, one must bear in mind that the correlation functions were obtained in different phase space regions $\Omega$ which enter into the integrations in (10): for the NA22-data, $\Omega$ is defined through the cuts $|y| < 2$, $k_\perp > 0$, whereas for the UA1-data one has $|y| < 3$, $k_\perp > 0.15\ GeV$. We have applied the fit parameters used for the description of the UA1-data (Fig. 1) to calculate $\tilde{C}_2(Q^2)$ in the kinematical region that corresponds to the NA22-data. The results, both for the static source and for the expanding source, are shown in Fig. 2. Obviously, they fail to reproduce the NA22-data. This implies that the observed difference in the widths of $\tilde{C}_2(Q^2)$ in the NA22- and in the UA1-experiment cannot be explained through the difference in the rapidity and transverse momentum cuts alone. Rather, it reflects a genuine dependence of the BEC on the CM energy $\sqrt{s}$ of the collision.

To describe the NA22-data, one needs to choose a set of parameter values different from the ones used to fit the UA1-data. The resultant space-time parameters are listed in Tables 1 and 2, and the corresponding correlation functions are compared to the

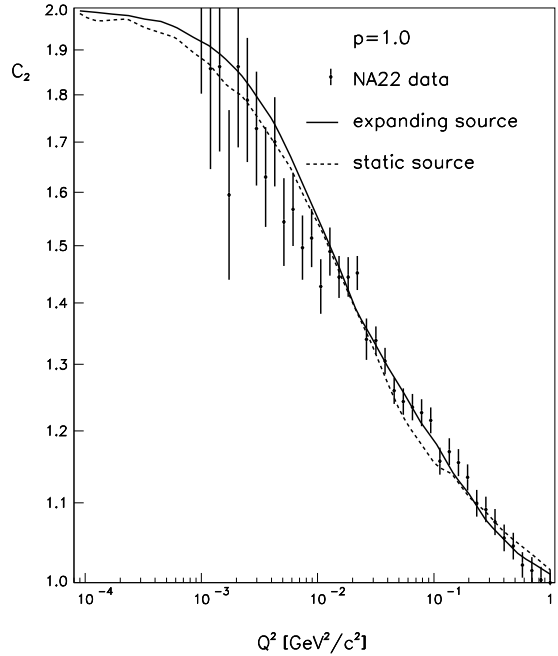

Figure 3: Best fits to the NA22 data

NA22-data in Fig. 3.

Up to this point, we have always assumed that the particles are emitted from a purely chaotic source. Before we can discuss the quality of the fits and the $s$-dependence of the quantum statistical parameters of the source, we have to address the question if the data allow any conclusions concerning the presence of a coherent component and its energy dependence.

The $\chi^2/d.o.f.$ for the best fits are plotted as a function of the chaoticity $p$ in Figs. 4a and 4b, for the UA1-data and the NA22-data, respectively, where the dashed curves correspond to the static source and the solid curves to the expanding source. According to the $\chi^2$-criterion, parameter sets that correspond to points above the straight line labeled 5% (1%) can be ruled out at the 5% (1%) level, i.e., the probability that the data are consistent with the model is below 5% (1%). At $\sqrt{s} = 630\ GeV$, the static source can be ruled out on the 5% level. On the other hand, the expanding source can fit the data with reasonable values of $\chi^2/d.o.f.$ for chaoticities in the range $0.9 \leq p \leq 1$. For the data at $\sqrt{s} = 22\ GeV$, both the static and the expanding source models are acceptable for a wide range of chaoticity parameters ($0.37 \leq p \leq 1$ for the static source and $0.31 \leq p \leq 1$ for the expanding source). As an example, we have listed the values of the fit parameters and $\chi^2/d.o.f.$ for a chaoticity of 0.6 at $\sqrt{s} = 22\ GeV$ in Tables 1 and 2. Note, however, that even at NA22-energies one might have some doubts about the static source model as the fits require lifetimes of

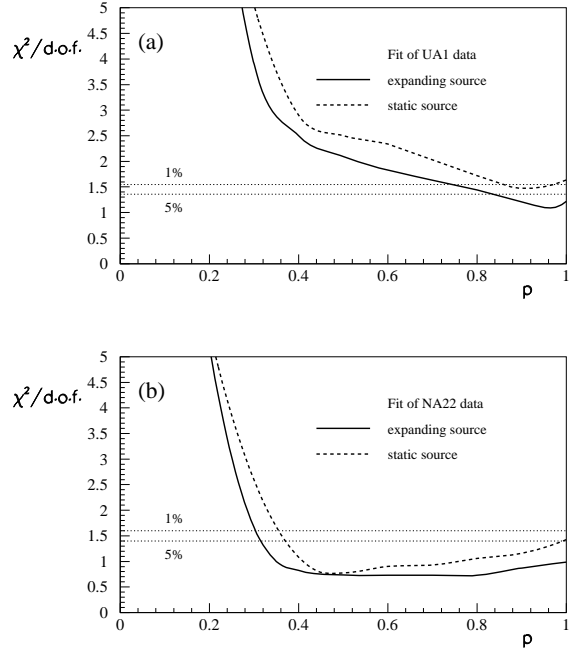

Figure 4: $\chi^2/d.o.f.$ for the best fits as a function of the chaoticity parameter $p$, (a) for the UA1-data and (b) for the NA22-data. As before, the dashed curves are for the static and the solid curves for the expanding source model. Points above the horizontal line labeled 5% (1%) correspond to chaoticity parameter values with a probability below 5% (1%) for the data to be consistent with the model

3 to 4 $fm$, i.e., values which are suspiciously large for hadronic collisions. Nevertheless, in principle one cannot exclude the possibility that particles are emitted from a static source at $\sqrt{s} = 22~GeV$ and from an expanding source at $\sqrt{s} = 630~GeV$. As was pointed out in [9], a sensitive test which would allow a clear distinction between these two types of sources is to measure $C_2$ as a function of the two variables $q_0^2$ and $\vec{q}^{\,2}$. There are no published data in these two variables at $\sqrt{s} = 22~GeV$ and at $\sqrt{s} = 630~GeV$, and experimentalists are urged to fill this gap.

Although a static source at $\sqrt{s} = 22~GeV$ cannot be completely ruled out at this point, in what follows we shall restrict our discussion of the energy dependence of the quantum statistical parameters to the expanding source model. As we have seen, the data at $\sqrt{s} = 22~GeV$ are consistent with chaoticity parameters in the range between 0.3 and 1. Fortunately, for the expanding source at 22 $GeV$ the values of the other fit parameters are not very sensitive to the value of $p$. As can be seen in Table 2, for $p = 1$ one has $R_{ch,\perp} = 0.65~fm$, $L_\eta = 0.90~fm$ and $\tau_0 = 1.90~fm$, while for $p = 0.6$ the values are $R_{ch,\perp} = 0.70~fm$, $L_\eta = 0.90~fm$ and $\tau_0 = 1.90~fm$. This enables us to draw conclusions about the $s$-dependence of $R_{ch,\perp}$, $L_\eta$ and $\tau_0$ although the NA22-data do not yield strong constraints on the value of $p$. Going from NA22- to UA1-energies, the longitudinal proper time $\tau_0$ is found to be almost unaffected (it changes from 1.9 $fm$ to 2 $fm$). On the other hand, the transverse radius $R_{ch,\perp}$ increases by about 40% (from $\sim 0.7\ fm$ to 1 $fm$) and the correlation length in longitudinal direction $L_\eta$ is reduced by more than 50% (it drops from 0.9 $fm$ to 0.4 $fm$). Concerning the chaoticity parameter one can conclude that it either increases with $s$ or remains approximately[4] constant. The correlation length in transverse direction $L_\perp$ decreases from the value of 0.8 $fm$ at 22 $GeV$ to 0.6 $fm$ at 630 $GeV$. If there is a coherent component, the transverse radius of the coherent source remains constant as a function of $s$, at a value of $R_{c,\perp} \sim 0.5\ fm$.

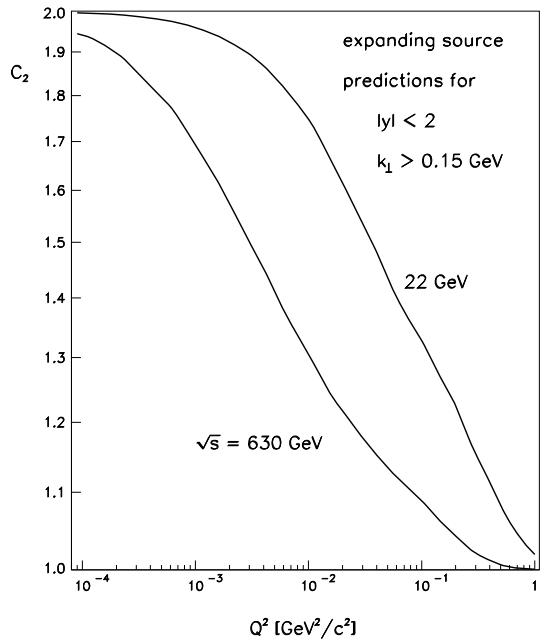

Figure 5: Predictions for $\tilde{C}_2(Q^2)$ in the momentum space region $|y| < 2$, $k_\perp > 0.15\ GeV$, for the expanding source at $\sqrt{s} = 22\ GeV$ and at $\sqrt{s} = 630\ GeV$. The fit parameter values are the same as in Figs. 1 and 3.

Of course it would be interesting to compare the BEC data at $\sqrt{s} = 22\ GeV$ and at $\sqrt{s} = 630\ GeV$ *in the same momentum space region*. The kinematical cuts are $|y| < 2$, $k_\perp > 0$ for the NA22-data and $|y| < 3$, $k_\perp > 0.15\ GeV$ for the UA1-data. We would therefore urge both experimental groups to re-analyze their data and present $\tilde{C}_2(Q^2)$ in the restricted phase space region $|y| < 2$, $k_\perp > 0.15\ GeV$. This would allow to explicitly see the $s$-dependence of the correlation function without having to take into account the effects of different phase space cuts. Fig. 5 shows our predictions for $\tilde{C}_2(Q^2)$ in the region $|y| < 2$, $k_\perp > 0.15\ GeV$ for the expanding source model, at $\sqrt{s} = 22\ GeV$ and at $\sqrt{s} = 630\ GeV$. The parameter values of $\tau_0$, $R_{ch,\perp}$, $L_\eta$ and $p = 1$ correspond to the fits to the data in Figs. 1 and 3. As can be seen by comparing Fig. 5 with Fig. 1, for the expanding source the restriction in rapidity (from $|y| < 3$ to $|y| < 2$) leaves the shape of the correlation function practically unchanged (cf. also [9]). On the other hand, a comparison of Fig. 3 and Fig. 5 shows that the cut in transverse momentum (from $k_\perp > 0$ to $k_\perp > 0.15\ GeV$) leads to a significant increase of the width of the correlation function. According to our prediction, the BEC function $\tilde{C}_2(Q^2)$ at $\sqrt{s} = 22\ GeV$ will be considerably broader than the one at $\sqrt{s} = 630\ GeV$, if the data at both energies are taken in the same momentum space region. This reflects the increase of the transverse radius and the decrease of the correlation lengths as a function of $s$.

---

[4]The data are also consistent with a slight decrease from $p \simeq 1$ at $\sqrt{s} = 22\ GeV$ to $p \simeq 0.9$ at $\sqrt{s} = 630\ GeV$, cf. Fig. 4.

Note that here we have not treated separately the contributions from the different resonance decays[5]. Thus the radii and lifetimes extracted from the BEC data and listed in Tables 1 and 2 must be regarded as effective radii and lifetimes in so far as they represent an average of the direct pion component and the components related to pions produced from resonance decays. We have also not attempted to correct for the effects of particle misidentification.

Our results concerning the $s$-dependence of the space-time parameters of the expanding source are in qualitative agreement with the experimental observation [12, 13, 14, 15] that at fixed $\sqrt{s}$ the width of the BEC function decreases with increasing rapidity density $dn/dy$. Since the average $dn/dy$ increases with $s$, the $dn/dy$-dependence of the BEC data suggests that the width of the correlation function decreases, which in the expanding source model is reflected in an increase of $R_{ch,\perp}$ and a decrease of $L_\eta$, $L_\perp$ with increasing CM energy. Unfortunately, the "effective radii" extracted from the data at fixed $dn/dy$ do not distinguish between components in beam direction and in the transverse plane. While the $s$-dependence of the space-time parameters agrees with what one would expect from the behaviour of the data as a function of $dn/dy$, this is not the case for the chaoticity parameter $p$. The data suggest a decrease of the $\lambda$-parameter with increasing $dn/dy$. While $\lambda$ does not agree with the quantum statistical chaoticity $p$), qualitatively a decrease of $\lambda$ translates into a decrease of $p$. On the other hand, in our present analysis of the data we have found that $p$ either increases or remains approximately constant as the CM energy increases from NA22- to UA1-energies. The resolution of this apparent discrepancy may be related to the fact that the extraction of a reliable value for the chaoticity parameter requires knowledge of the behaviour of the BEC function near the origin, i.e., at small momentum differences. It would therefore be useful also from this point of view if the experimentalists could measure $\tilde{C}_2(Q^2)$ at fixed $dn/dy$ down to values of $Q \simeq 30\ MeV$. The analysis of such data could help to determine the dependence of the quantum statistical parameters on the *two* variables $dn/dy$ and $s$.

---

[5] A formalism which explicitly takes into account the effect of resonance decays on BEC is presented in ref. [11].

In ref. [16], the dependence of the normalized factorial moments of the multiplicity distribution on the width of the rapidity interval was analysed for NA22- and for UA1-data. It was found that both the chaoticity parameter and the correlation length in rapidity $\xi$ increase with $s$. As an increase of $\xi$ corresponds to an increase of $L_\eta$ in the present formalism, the $s$-dependence of $\xi$ obtained in [16] would appear to be in contradiction with the $s$-dependence of $L_\eta$ obtained here. However, a note of caution is in order. Our present results cannot be compared to those of ref. [16] in a straightforward manner for a number of reasons. Firstly, the calculations in [16] were based on an ansatz for the correlation function in one variable only (rapidity). Secondly, at that time the factorial moments at Collider energies were known only for charged particles. Finally, and most importantly, in Ref. [16] it was assumed that only short-range correlations (SRC) contribute to the correlation function. However, recently it has been found [1] that at Collider energies there is a significant contribution due to long-range correlations (LRC), and that LRC increase strongly with $s$. Indeed it appears that in ref. [16] the strong increase of fluctuations due to the increase of LRC was attributed to an increase of SRC and hence, an increase of $\xi$ with the CM energy.

On the theoretical side, there exist surprisingly few predictions concerning the dependence of BEC on the CM energy $\sqrt{s}$. While the Lund model [17] apparently has no $s$ dependence for BEC a decrease of the width of the BEC function in longitudinal direction, $C_2(q_\parallel)$, with increasing $s$ was predicted by Bowler in the context of a string model[18]. A similar behaviour is also expected in a hydrodynamic scenario: for a longitudinally expanding source, the width of $C_2(q_\parallel)$ is approximately inversely proportional to the freeze-out proper time $\tau_f$ [19]. If the formation time $\tau_0$ of the thermalized matter is assumed to be $s$-independent, $\tau_f$ increases with the initial energy density which in turn increases with $s$. Consequently, the width of $C_2(q_\parallel)$ decreases with increasing CM energy.

An increase of the transverse radius with $s$ can be understood in the context of a simple geometric description of multiparticle production. From $\sqrt{s} = 22\ GeV$ to $630\ GeV$ the total $pp$ ($p\bar{p}$) cross section $\sigma_{tot}$ increases by about 60% (cf. the parametrization of $\sigma_{tot}(s)$ in ref. [20]). If one assumes that the transverse source radius scales approximately like $\sigma_{tot}^{1/2}$ one would predict an increase of $R_\perp$ by $\sim 30\%$. This is to be compared with the value of $\sim 40\%$ found in our analysis based on the model of a longitudinally expanding source[6]. For a hydrodynamically expanding source the $s$-dependence of the longitudinal and transverse flow components may also contribute to the increase of the transverse radius. As was shown in ref. [21], for reasonable values of the freeze-out temperature transverse flow can lead to a *decrease* of the transverse source radius extracted from BEC data (cf. Fig. 3 in [21]). At higher energies, the longitudinal expansion dominates and transverse flow becomes less important (cf. ref. [22]). Therefore, the transverse source radius can be expected to increase with the CM energy.

---

[6]Indeed, the fact that an increase of $R_\perp$ is to be expected from simple phenomenological considerations can be regarded as further support for the expanding source model and evidence against the validity of the static source model. For the case of a static source at $\sqrt{s} = 22\ GeV$, our analysis of the data would imply that the transverse radius takes approximately the same value at $22\ GeV$ and at $630\ GeV$ (cf. Tables 1 and 2).

The fact that the observed change with energy of source geometry is in agreement with expectations derived from independent considerations proves that the theory of BEC with its space-time formulation [4] has come out of age and gives confidence for its use in applications like quark-gluon plasma search. This conclusion is reinforced by the good agreement of "intermittency" data with results obtained from Bose-Einstein correlations [8, 9].

This work was supported by the Federal Minister of Research and Technology under contract 06MR731 and the Deutsche Forschungsgemeinschaft. We are indebted to B. Buschbeck (UA1 Collaboration), W. Kittel (NA22 Collaboration) and U. Ornik for many instructive discussions.